\documentclass[journal]{IEEEtran}
\usepackage{amsmath,amsfonts}
\usepackage{algorithmic}
\usepackage{algorithm}
\usepackage{array}
\usepackage[caption=false,font=footnotesize,labelfont=rm,textfont=rm]{subfig}
\usepackage{textcomp}
\usepackage{stfloats}
\usepackage{url}
\usepackage{verbatim}
\usepackage{graphicx}
\usepackage{cite}
\usepackage{booktabs}
\usepackage{threeparttable}
\usepackage{bm}                    
\usepackage{amssymb}
\usepackage{hyperref}
\usepackage{orcidlink}
\usepackage{lineno}
\usepackage{tabularray}
\usepackage{multirow}
\usepackage{makecell}
\usepackage{xcolor}
\usepackage{microtype}
\usepackage{caption}
\usepackage{siunitx}
\usepackage{enumitem}

\graphicspath{{./figure}{./bio}}

\begin{document}


\title{
    A Neural Network-Based Real-time Casing Collar Recognition System for Downhole Instruments
}

\author{
    Si-Yu~Xiao      \orcidlink{0009-0006-3095-3242},
    Xin-Di~Zhao     \orcidlink{0009-0007-4304-001X},
    Xiang-Zhan~Wang \orcidlink{0000-0003-1898-4629},
    Tian-Hao~Mao    \orcidlink{0009-006-1898-1080},
    Ying-Kai~Liao   \orcidlink{0009-0002-3758-3679},
    Xing-Yu~Liao    \orcidlink{0009-0009-9779-5869},\linebreak
    Yu-Qiao~Chen    \orcidlink{0009-0000-7241-8850},
    Jun-Jie~Wang    \orcidlink{0000-0001-7183-422X},
    Shuang~Liu      \orcidlink{0000-0002-0587-4415},
    Tu-Pei~Chen     \orcidlink{0000-0002-1098-9575} and
    Yang~Liu        \orcidlink{0000-0003-0615-7036}\IEEEauthorrefmark{1}
\thanks{Si-Yu Xiao, Xiang-Zhan Wang, Tian-Hao Mao, Ying-Kai Liao, Xing-Yu Liao, Yu-Qiao Chen, Jun-Jie Wang, Shuang Liu and Yang Liu are with the State Key Laboratory of Electronic Thin Films and Integrated Devices, University of Electronic Science and Technology of China, Chengdu 611731, China.}
\thanks{Xin-Di Zhao is with the Southwest Branch of China National Petroleum Corporation Logging Co., Ltd., Chongqing 401100, China.}
\thanks{Tu-Pei Chen is with the School of Electrical and Electronic Engineering, Nanyang Technological University, Singapore 639798.}
\thanks{This work is supported by NSFC under project No.~62404034 and 62404033. This work is also supported by China National Petroleum Corporation Logging Co., Ltd. (CNLC) under project No.~CNLC2023-7A01.}
\thanks{\IEEEauthorrefmark{1}Corresponding author.}
}


\maketitle

\begin{abstract}
Casing collar locator (CCL) measurements are widely used as reliable depth markers for positioning downhole instruments in cased-hole operations, enabling accurate depth control for operations such as perforation. However, autonomous collar recognition in downhole environments remains challenging because CCL signals are often corrupted by toolstring- or casing-induced magnetic interference, while stringent size and power budgets limit the use of computationally intensive algorithms and specific operations require real-time, in-situ processing.
To address these constraints, we propose Collar Recognition Nets (CRNs), a family of domain-specific lightweight 1-D convolutional neural networks for collar signature recognition from streaming CCL waveforms.
With depthwise separable convolutions and input pooling, CRNs optimize efficiency without sacrificing accuracy. Our most compact model achieves an F1-score of 0.972 on field data with only 1,985~parameters and 8,208~MACs, and deployed on an ARM Cortex-M7 based embedded system using TensorFlow Lite for Microcontrollers (TFLM) library, the model demonstrates a throughput of 1,000 inference per second and \SI{343.2}{\micro\second} latency, confirming the feasibility of robust, autonomous, and real-time collar recognition under stringent downhole constraints.
\end{abstract}

\begin{IEEEkeywords}
Artificial Intelligence, Casing collar locator, Deep Learning, Downhole Instrument, Edge Computing System, Industrial Automation, Pattern Recognition, Signal Processing
\end{IEEEkeywords}

\section{Introduction}

\begin{figure*}[!htbp]
    \centering
    \includegraphics[width=1\linewidth]{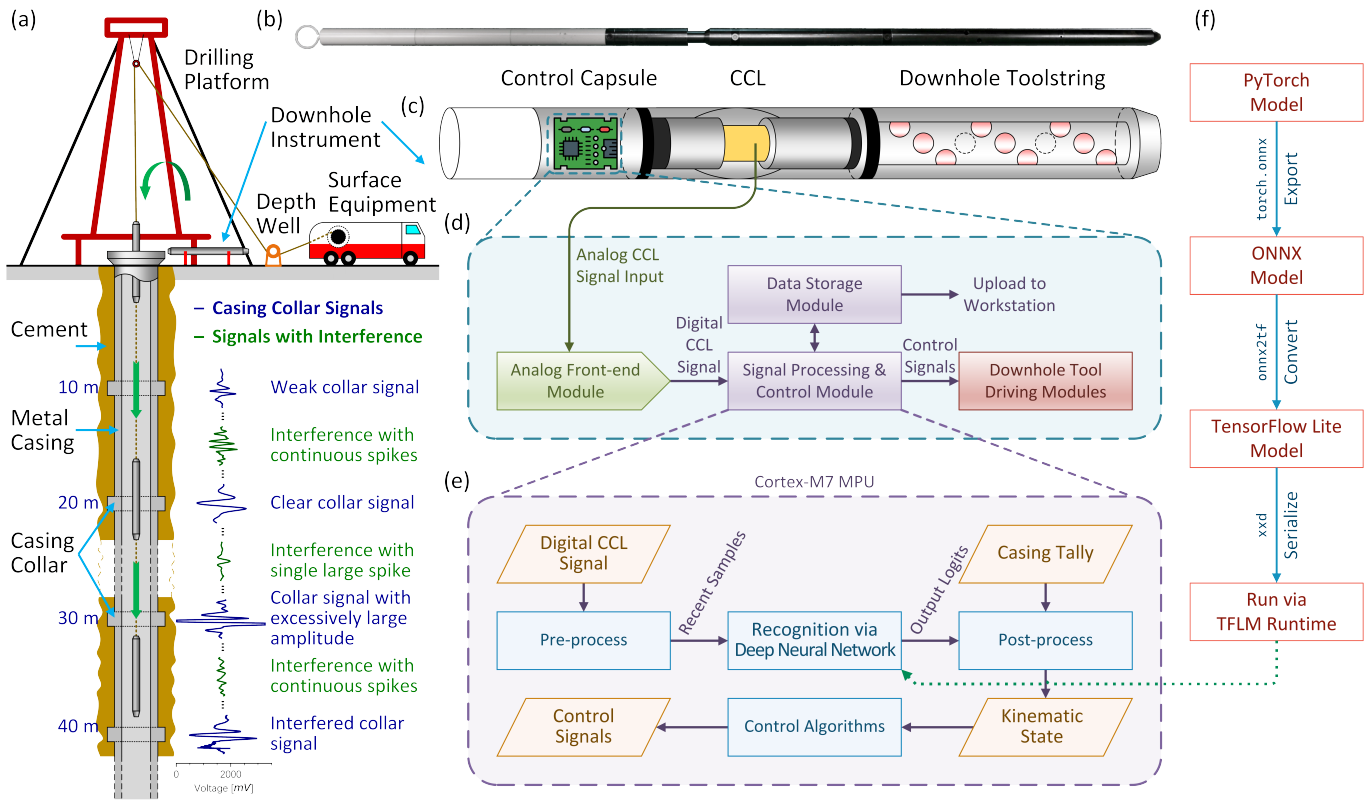}
    \caption{
        \textbf{(a)}~Schematic cross-section of a typical oil and gas well structure. Representative casing collar signatures derived from magnetic response are illustrated in dark blue near the corresponding casing collars, while typical interference signals are illustrated in dark green. Adapted from \cite{xiao2025realization,xiao2025dataaugmented};
        \textbf{(b)}~Photograph of an perforating gun, exemplifying a typical downhole instrument assembly used in oil and gas wells;
        \textbf{(c)}~Schematic diagram of the internal structure of a downhole instrument; the battery is omitted for clarity;
        \textbf{(d)}~Block diagram of the collar recognition system within the control capsule;
        \textbf{(e)}~Progress flow diagram for casing collar recognition utilizing neural network and casing tally;
        \textbf{(f)}~Deployment workflow for neural network models, illustrating the transition from a PyTorch model to an executable program.
    }
    \label{fig1}
\end{figure*}

\IEEEPARstart{I}{n} the exploration and production of oil and gas resources, the accurate positioning of downhole instruments remains a challenging yet critical task \cite{harris1966effect,raman2024data}, as it directly influences maximum productivity and operational safety \cite{deffenbaugh2017untethered}.
The detection of casing collars, which serve as depth markers along the steel casing string, by casing collar locators (CCLs), is a predominant method for estimating the depth of downhole instruments due to its cost-effectiveness, efficiency, and high reliability \cite{alvarez2018theory,gidado2023well,mijarez2014hpht}, as illustrated in Fig.~\ref{fig1}(a). A CCL is a magnetic sensor typically integrated into the downhole toolstring, comprising a coil positioned between two permanent magnets.
As the CCL traverses a casing collar \cite{li2013casing,seren2022miniaturized}, the magnetic flux lines concentrate within the increased metal mass of the casing collar \cite{li2013casing,mijarez2014hpht,raman2024data} and this variation in the magnetic field induces a voltage pulse in the coil \cite{li2013casing,alvarez2018theory,gidado2023well}, referred to as a ``CCL response'' or ``collar (magnetic) signature''. This characteristic magnetic response typically exhibits a bipolar signature \cite{mijarez2014hpht,seren2022miniaturized,song2023finite}, as illustrated by the dark blue traces in Fig.~\ref{fig1}(a).
By correlating these collar signatures with the ``casing tally'' (reference depths derived from well completion records) \cite{seren2022miniaturized,seren2025probabilistic}, the precise position of the downhole instrument can be determined \cite{li2013casing,alvarez2018theory,seren2022miniaturized,gidado2023well}. For instance, in perforation operations, this positioning allows the perforating gun to target specific intervals with high accuracy. However, the running-in-hole (RIH) speed significantly influences the morphology of the CCL response, posing a substantial challenge to the development of automatic recognition methods.

Additionally, while the surface wheel measurement (SWM) method is a common and straightforward approach \cite{alvarez2018theory,raman2024data}, it is practically unreliable due to the inherent elasticity of long metallic wirelines \cite{raman2024data}, providing only a rough estimation of the downhole instrument position. Furthermore, SWM is inapplicable to emerging operations such as wireless perforating \cite{entchev2011autonomous,seren2022miniaturized}. In such scenarios, utilizing CCL provides a more advanced approach to accurate downhole positioning.

In standard operations, the recognition of collar signatures from CCL logs is typically performed at the surface \cite{seren2022miniaturized}, relying on data transmitted via wirelines exceeding thousands of meters \cite{mijarez2014hpht,alvarez2018theory}. However, this approach encounters significant practical challenges:
\begin{enumerate}[label=(\arabic*)]
\item The CCL signal is frequently contaminated by extraneous magnetic sources, such as the casing wall and the metallic downhole toolstring \cite{alvarez2018theory,zeng2022cclsignal}. As illustrated in Fig.~\ref{fig1}(a), certain interference waveforms (green) exhibit high similarity to authentic collar signatures (blue) and occupy overlapping frequency bands, thereby complicating the discrimination between the genuine CCL response and interference \cite{wang2006application,alvarez2018theory}.
\item Extended wirelines introduce substantial signal attenuation and noise, impeding the reliable recognition of collar signatures \cite{brown1990effects,mijarez2014hpht,zhao2021highspeed}. Notably, stochastic noise contributes minimally to the overall interference profile. Consequently, conventional filtering techniques are often ineffective in suppressing the dominant, complex interference components.
\item Manual identification of collar signatures at the surface is inefficient and prone to error \cite{alvarez2018theory,zeng2022cclsignal}. Moreover, the reliance on human interpretation and surface equipment incurs substantial operational costs \cite{deffenbaugh2017untethered,raman2024data}. Furthermore, advanced operations such as wireless perforating \cite{entchev2011autonomous,seren2023magnetic}, pump-down perforation (PDP), and plug-and-perf (P\&P) \cite{lu2012oilandgasfield} require real-time collar recognition \cite{ahmed2021case}, thereby compounding the processing challenges.
\item Due to the restricted diameter of the wellbore, the available space and power budget within the downhole instrument are severely limited \cite{zhao2021highspeed}. These constraints preclude the integration of conventional high-performance, power-intensive computing hardware \cite{seren2023magnetic}.
\end{enumerate}
To enhance the automation of CCL-based downhole operations and address these limitations, the development of automatic, in-situ, and real-time CCL signal processing using lightweight algorithms embedded within downhole instruments is highly desirable.

Traditional methods include thresholding techniques \cite{wang2012collardepth,cong2022perforating}, time-domain filtering \cite{li2020application}, frequency-domain analysis \cite{li2013casing,li2010approach}, and physical plausibility analysis \cite{xiao2025realization,alvarez2018theory}. However, these approaches exhibit limited generalizability \cite{li2020application,zeng2022cclsignal,yang2025leak}, making it challenging to recognize collar signatures in the presence of complex and variable interference \cite{xiao2025realization}. While the algorithm based on multi-hypothesis localization and Bayesian filtering demonstrate superior performance \cite{seren2025probabilistic}, it necessitates the use of multi-modal sensors.

Driven by rapid advancements and increasing accessibility, deep neural networks (DNNs) have emerged as a promising solution for complex challenges in the oil and gas industry. Key applications range from data interpretation \cite{noh2021deep,brazell2019machine} and imaging \cite{viggen2025improving} to operational planning \cite{elhadidy2025optimizing} and drilling rate evaluation \cite{liu2025research}. Specifically, in the domain of collar signature recognition, architectures such as convolutional neural networks (CNNs) \cite{zhao2022detection,zhang2024yolo,raman2024data,jing2025identification,torrescaceres2024automated}, recurrent neural networks (RNNs) \cite{raman2024data,jing2025identification,lu2025digital}, and residual neural networks (ResNets) \cite{yan2024automatic} have yielded promising results. However, the high computational overhead of these models often exceeds the processing capabilities of downhole edge devices. Consequently, optimizing model architecture is essential to ensure compatibility with these hardware constraints.

Over the past decade, neural network theory has evolved significantly. Causal and temporal convolutions provide a theoretical foundation for applying convolution operators to signal processing \cite{bai2018empirical,oord2016wavenet,van2016conditional}. To enhance computational efficiency, architectures such as MobileNets \cite{howard2017mobilenets,sandler2018mobilenetv2,howard2019searching} and Network in Network (NIN) \cite{lin2013network} have popularized critical techniques. These include the use of consecutive small convolutional kernels \cite{simonyan2015very}, depthwise separable convolutions (DSCs or DW-Convs) combined with pointwise convolutions (PW-Convs) \cite{chollet2017xception}, and global average pooling \cite{szegedy2016rethinking}. Additionally, batch normalization \cite{ioffe2015batch} and dropout \cite{srivastava2014dropout,srivastava2013improving,bouthillier2015dropout,tompson2015efficient,ghiasi2018dropblock} have been established as standard regularization techniques to mitigate overfitting and improve generalization. Furthermore, the attention mechanism has revolutionized feature extraction in deep learning \cite{hu2018squeeze,vaswani2017attention,dosovitskiy2021an}. Finally, the emergence of tiny machine learning (TinyML) has facilitated the deployment of neural networks on devices with severe memory and power constraints \cite{wang2017timeseries,zhang2018helloedge,lin2020mcunet,lin2021mcunetv2,kim2021broadcasted,lin2022ondevice}.

To address the challenge of precise, real-time collar recognition from interference-contaminated CCL signals, specifically in wireless downhole instruments with limited resources, this paper proposes Collar Recognition Nets (CRNs). These are three lightweight 1D-CNN architectures tailored for casing collar recognition. Unlike prior approaches that prioritize accuracy at the expense of complexity, we explicitly balance computational efficiency with recognition performance. By leveraging depthwise separable convolutions and input pooling, we reduce the computational cost of our most compact variant, CRN-3, to just 1,985 parameters and 8,208 Multiply-Accumulate operations (MACs), while maintaining an F1 score of 0.972. This model is successfully deployed on a custom downhole instrument powered by an ARM Cortex-M7 microprocessor unit (MPU). We validate the performance of both the proposed models and the integrated system using field data. Notably, the computational complexity of this domain-specific model is orders of magnitude lower than generic TinyML baselines, such as the minimized DS-CNN-small (\textasciitilde 200K MACs) \cite{zhang2018helloedge} and MobileNetV3-small-0.75 (2.4M parameters and 44M MACs) \cite{howard2019searching}.

The main contributions of this paper are summarized as follows:
\begin{enumerate}[label=(\arabic*)]
\item We propose CRNs, a family of domain-specific neural networks tailored for casing collar recognition. By leveraging depthwise separable convolutions and input pooling, we balance computational complexity with recognition performance, achieving a model complexity significantly lower than standard TinyML baselines while maintaining high recognition accuracy.
\item We demonstrate the feasibility of deploying the lightweight CRN-3 model on an ARM Cortex-M7 MPU using the TensorFlow Lite for Microcontrollers (TFLM) \cite{david2021tensorflow} runtime. The system achieves a throughput of 1,000 inferences per second (IPS), enabling robust real-time processing.
\item We validate the proposed system using field data, demonstrating its capability to accurately identify collar signatures under strict in-situ operational constraints.
\end{enumerate}

\section{Methodology}

\subsection{Problem Formulation}

The primary objective of this study is to develop a collar recognition system embedded within a downhole toolstring to enable in-situ, real-time identification. Subsequently, these recognition results are correlated with the casing tally to refine the positional estimation of the downhole instrument.

Preliminary analysis reveals that the raw signal acquired from the CCL is an analog waveform comprising collar signatures, interference signals, and noise, as illustrated in Fig.~\ref{fig1}(a). Both casing collar signatures and interference signals typically occupy the frequency band of \SIrange{10}{40}{\hertz}, whereas noise is predominantly concentrated in the high-frequency spectrum.
Furthermore, based on the characterization of CCL signals in \cite{song2023finite}, the temporal width of the collar signatures is inversely proportional to the logging speed, negatively correlated with the CCL coil length, and independent of the casing collar length.

In conventional processing, the CCL signal is represented as a continuous waveform. Identifying collar signatures relies on manual interpretation by experts to distinguish genuine signatures from interference, and is a labor-intensive, subjective, and time-consuming process \cite{raman2024data} . In this work, recognition is automated using a neural network that treats the CCL waveform as a one-dimensional sequence. Upon recognition of collar signatures, the corresponding depth is determined by correlating these detection events with the known casing tally derived from well completion records. Consequently, the kinematic state of the downhole toolstring is inferred via analysis of the timestamps and depths of the identified collars \cite{xiao2025dataaugmented} . Given the constraints of wireless perforating operations, this entire workflow must be executed in real-time and in situ within the downhole instrument.

\subsection{System Configuration}

The configuration of the downhole toolstring is illustrated in Fig.~\ref{fig1}(b--c). This assembly primarily comprises a control capsule, a CCL sensor, and downhole payloads. The collar recognition system is integrated within the control capsule. Representing an advancement over the embedded design originally proposed in \cite{xiao2025realization}, the system architecture is detailed in Fig.~\ref{fig1}(d). The core of the recognition unit is a microprocessor unit (MPU), as shown in Fig.~\ref{fig1}(e). This component offers a compact form factor and possesses limited computational resources compared to the high-performance hardware typically deployed in surface workstations \cite{lin2020mcunet}.

To withstand high-temperature and high-pressure (HPHT) downhole conditions \cite{mijarez2014hpht}, the control capsule utilizes a thermally insulated and pressure-resistant housing. Furthermore, to ensure system reliability and stability, the internal electronic components are selected from automotive-grade specifications, supporting an operating temperature range of \SIrange{-40}{125}{\celsius}.

\subsection{Signal Processing}

\subsubsection{Acquisition of CCL Signal}

\begin{figure}[!t]
    \centering
    \includegraphics[width=0.825\linewidth]{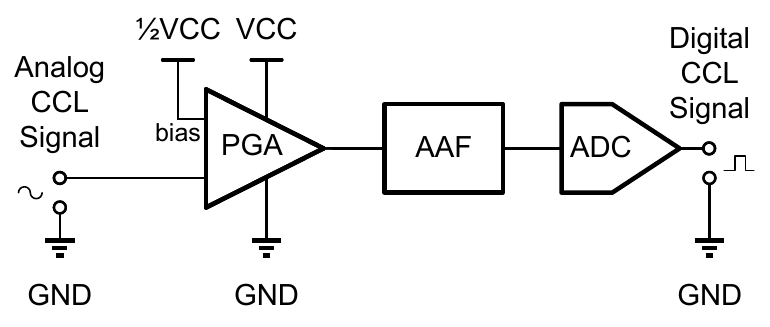}
    \caption{
        Schematic diagram of the analog front-end (AFE) module within the collar recognition system, comprising a programmable gain amplifier (PGA), an anti-aliasing filter (AAF), and an analog-to-digital converter (ADC).
    }
    \label{fig2}
\end{figure}

The raw analog CCL signal is acquired via the analog front-end (AFE) module, as illustrated in Fig.~\ref{fig1}(d). As detailed in Fig.~\ref{fig2}, the AFE comprises a programmable gain amplifier (PGA), an anti-aliasing filter (AAF), and an analog-to-digital converter (ADC). The PGA adjusts the signal amplitude to the required dynamic range. The AAF functions as a low-pass filter to attenuate high-frequency noise. The ADC digitizes the raw CCL signal at a sampling rate of \SI{1}{\kilo\hertz} with 16-bit resolution, and the resulting digital stream is transmitted to the MPU for subsequent processing. Given the low-frequency characteristics of the signal encountered in practical applications, this streamlined hardware architecture proves sufficient.

\subsubsection{Pre-processing of CCL Signal}

The integer values of the digitized CCL signal are initially recorded in data storage. Subsequently, these values are normalized to a floating-point format using an empirical formula, approximating a standard normal distribution (zero mean, unit variance). A sliding window buffer retains the most recent 160 samples, which are then fed into the neural network as the input sequence.

\subsubsection{Processing by Deep Neural Network}

The neural network processes the buffered sequence to estimate the probability of a collar signature at the current time step. This probability is derived by applying a sigmoid activation function to the network's output logit and is subsequently stored. Over time, a temporal probability map is constructed, where each point represents the likelihood of a collar occurrence {at that specific instance.}

\subsubsection{Post-processing of Recognition Results}

Based on the definition of the probability map and the label processing method detailed in Section~\ref{sec:network}, the occurrence of a collar signature manifests as a distinct pulse in the probability map, as illustrated in Fig.~\ref{fig4}. Consequently, a fixed threshold facilitates detection, generating a binary sequence. In this study, the threshold is set to 0.5; this value aligns with standard probabilistic definitions and balances the trade-off between sensitivity and specificity.
Since the raw probability map may exhibit fluctuations or spikes, the binary sequence undergoes a smoothing process involving expansion and contraction, implemented via a running average filter. This smoothed sequence is subsequently utilized to calculate the temporal centroid of the collar signature.
A similar algorithm is employed in \cite{xiao2025dataaugmented} for centroid estimation. Finally, the timestamps of the identified signatures are recorded in storage as recognition results.

From the recognition results, the timestamps and depths of the identified collars are determined. The three most recent timestamp-depth pairs serve as boundary conditions in kinematic equations to estimate the kinematic state and current depth of the downhole toolstring. This information is subsequently utilized by control algorithms to generate signals for operating the downhole payloads.

\subsection{Network Structure and Training}\label{sec:network}

Due to the limited computational resources of MPUs, the depth, feature size, and parameters of the proposed model must be strictly controlled. We leverage Temporal Convolutional Network (TCN) \cite{bai2018empirical} inspired design principles within a MobileNet framework to maintain efficiency. Self-attention mechanisms are omitted to avoid excessive computational costs, especially considering the limited computational capacity of the downhole instrument \cite{hu2018squeeze,dosovitskiy2021an,xiao2025dataaugmented}.

\begin{figure}[!tb]
    \centering
    \includegraphics[width=0.8\linewidth]{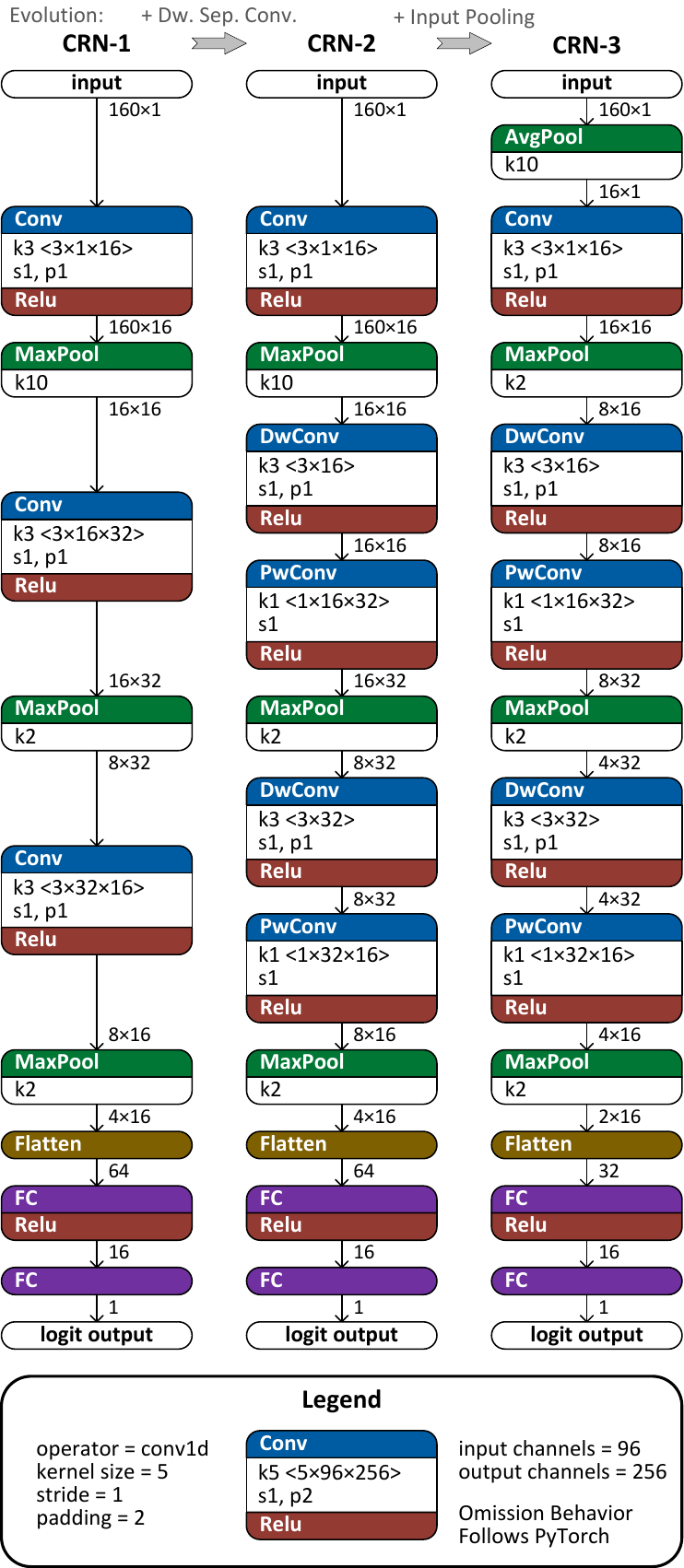}
    \caption{
        Network architectures of the Collar Recognition Nets (CRNs) proposed in this work. For clarity, the batch normalization and dropout layers following each convolutional layer or fully connected layer are omitted.
    }
    \label{fig3}
\end{figure}

The architectures of proposed models, designated as Collar Recognition Nets (CRNs), are illustrated in Fig.~\ref{fig3}. The model input is a sequence comprising the most recent 160 sample points (equivalent to \SI{160}{\milli\second}) from the pre-processed CCL signal. Statistical analysis of the existing dataset, combined with the theoretical characterization of the CCL signal, indicates that the selected window length is sufficient for capturing signature features. As exemplified in Fig.~\ref{fig4}(a), the collar signature exhibits a duration of approximately \SI{100}{\milli\second}.
The backbone employs standard blocks (convolution, batch normalization (BN) \cite{ioffe2015batch}, spatial dropout \cite{tompson2015efficient}, and activation), while the head consists of fully connected (FC), activation, and dropout layers. The output logit indicates the probability of a collar signature.
Although CRN-1 provides a functional baseline with sufficient parameter capacity to reach maximum performance, it is not optimized for resource-constrained deployment. To improve efficiency, CRN-2 utilizes depthwise separable convolutions, which cut the convolutional computational cost to approximately 40\% of the baseline \cite{howard2017mobilenets}. Building on this, CRN-3 introduces an initial average pooling layer with a kernel size of 10 to further reduce the computational load with negligible impact on accuracy, as illustrated in Fig.~\ref{fig3} and Table~\ref{tab2}.
Given that the CCL signal sampling rate is \SI{1}{\kilo\hertz}, the equivalent sampling rate is reduced to \SI{100}{\hertz}. Since the signal bandwidth is limited to \SI{40}{\hertz}, this rate exceeds the required Nyquist rate of \SI{80}{\hertz}, preventing information loss due to aliasing.

The CCL logs employed in this work were acquired from field operations in Sichuan Province, China, ensuring that the results are representative of realistic downhole conditions. The dataset was partitioned into training and validation sets using a 3:1 ratio based on the CCL logs, yielding 288 and 50 collar signature samples, respectively. Each sample is associated with a manually annotated ground truth label representing the centroid timestamp of the collar signature.
The initial dataset size is constrained by the limited number of available wells and operational restrictions. To enhance training efficiency and data diversity, we employ a suite of augmentation techniques: label distribution smoothing (LDS), random cropping, label smoothing regularization (LSR), time scaling, and multiple sampling \cite{xiao2025dataaugmented}. Notably, multiple sampling involves stochastically processing a single sample-label pair multiple times (e.g., via random cropping) to generate a set of distinct augmented training instances. This strategy increases the volume of training data while mitigating the risk of overfitting.
Utilizing these strategies, the training and validation sets are expanded by a factor of 20, reaching 5,760 and 1,000 samples, respectively. Furthermore, collar labels are converted into temporal probability maps by LDS, as illustrated in Fig.~\ref{fig4}(a), to provide more informative training targets \cite{xiao2025dataaugmented}.

The model is implemented within the PyTorch framework and trained using the Binary cross-entropy (BCE) loss function and the Adam optimizer. Both training and validation processes were conducted offline on a workstation.

\subsection{Deployment of Networks}

The system is powered by an ARM Cortex-M7 MPU, which features a double-precision floating point unit (FPU), single instruction multiple data (SIMD) capabilities, and an L1-cache. Experimental memory-bound benchmarks indicate that the processor achieves a throughput of approximately 70 MFLOPS at a clock frequency of \SI{550}{\mega\hertz} under optimal conditions.

The parameters counts and computational costs of the evaluated models are summarized in Table~\ref{tab2}. Specifically, the CRN-3 model comprises 1,985 parameters and requires 8,208 MACs per inference. The deployment of the CRN-3 model on this MPU is computationally feasible, supporting continuous \SI{1}{\kilo\hertz} inference to match the sensor's sampling rate. TensorFlow Lite for Microcontrollers (TFLM) \cite{david2021tensorflow} serves as the runtime environment for executing neural network models on microcontroller devices. TFLM provides optimized implementations for the Cortex-M architecture, leveraging the hardware-accelerated FPU and SIMD instructions. The CRN-3 model is converted into a deployable format using TFLM tools and integrated into the recognition firmware, as illustrated in Fig.~\ref{fig1}(e)--(f).

\section{Validation and Discussion}

\newcommand{\cdash}{        
    \multicolumn{1}{c}{--}
}

\begin{table}[!b]
\caption{
    Performance Metrics of the Proposed Models
}
\label{tab1}

\centering
\begin{minipage}{\linewidth}
\captionsetup{font=large, justification=raggedright, singlelinecheck=false}
\setlength{\tabcolsep}{5pt}

\resizebox{\linewidth}{!}{
\begin{tabular}{l rrrr rrrr}
\toprule[1pt]

\multicolumn{1}{c}{\multirow{2}{*}{\textbf{Model}}} &
\multicolumn{4}{c}{\textbf{Confusion Matrix}} &
\multicolumn{4}{c}{\textbf{Classification Metrics}} \\

\cmidrule[1pt](lr){2-5} \cmidrule[1pt](lr){6-9}

&
\multicolumn{1}{c}{\textbf{TP}} &
\multicolumn{1}{c}{\textbf{FP}} &
\multicolumn{1}{c}{\textbf{FN}} &
\multicolumn{1}{c}{\textbf{TN}} &
\multicolumn{1}{c}{\textbf{Acc}} &
\multicolumn{1}{c}{\textbf{P}} &
\multicolumn{1}{c}{\textbf{R}} &
\multicolumn{1}{c}{\textbf{F1}} \\

\midrule[1pt]

CRN-1 & 127 & 0 & 2 & 0 & 98.4\% & 100.0\% & 98.4\%  & 99.2\% \\
\midrule
CRN-2 & 126 & 0 & 3 & 0 & 97.7\% & 100.0\% & 97.7\%  & 98.8\% \\
\midrule
CRN-3 & 122 & 0 & 7 & 0 & 94.6\% & 100.0\% & 94.6\%  & 97.2\% \\

\midrule[1pt]

TAN   & 129 & 3 & 0 & 0 & 97.7\% & 97.7\%  & 100.0\% & 98.9\% \\
\midrule
MAN   & 127 & 3 & 2 & 0 & 96.2\% & 97.7\%  & 98.4\%  & 98.1\% \\

\bottomrule[1pt]
\end{tabular}
}
\vspace{0.5ex}
\noindent{\footnotesize\\
``TP'', ``FP'', ``FN'', ``TN'' denote true positive, false positive, false negative, and true negative, respectively; ``Acc'', ``P'', ``R'', ``F1'' denote accuracy, precision, recall, and F1 score, respectively.
}
\end{minipage}
\end{table}

\let\cdash\undefined

\newcommand{\cdash}{        
    \multicolumn{1}{c}{--}
}

\begin{table*}[!b]
\centering
\captionsetup{font=large, justification=raggedright, singlelinecheck=false}
\caption{
    Comparison of Network Capacity and Performance with Existing Works
}
\label{tab2}

\begin{minipage}{0.95\linewidth}
\setlength{\tabcolsep}{5pt}

\resizebox{\linewidth}{!}{
\begin{tabular}{r@{\hspace{3pt}}l rr rrrr l rc}
\toprule[1pt]

\multicolumn{2}{c}{\multirow{2}{*}{\textbf{Method}}} &
\multicolumn{2}{c}{\textbf{Network Capacity}} &
\multicolumn{4}{c}{\textbf{Classification Metrics}} &
\multicolumn{1}{c}{\multirow{2}{*}{\textbf{Remarks}}} &
\multicolumn{1}{c}{\multirow{2}{*}{\makecell[c]{\textbf{Inference}\\\textbf{Latency}}}} &
\multicolumn{1}{c}{\multirow{2}{*}{\makecell[c]{\textbf{In-situ,}\\\textbf{Real-time}}}} \\

\cmidrule[1pt](lr){3-4} \cmidrule[1pt](lr){5-8}

& &
\multicolumn{1}{c}{\textbf{Params}} &
\multicolumn{1}{c}{\textbf{MACs}} &
\multicolumn{1}{c}{\textbf{Acc}} &
\multicolumn{1}{c}{\textbf{P}} &
\multicolumn{1}{c}{\textbf{R}} &
\multicolumn{1}{c}{\textbf{F1}} &
& & \\

\midrule[1pt]

& CRN-1 & 4305 & 45584 & 98.4\% & 100.0\% & 98.4\% & 99.2\% & CNN + Data Augmentation                        & {\SI{1465.0}{\micro\second}} & \\
\midrule
& CRN-2 & 2497 & 22544 & 97.7\% & 100.0\% & 97.7\% & 98.8\% & CNN + Data Augmentation + DSC                  & {\SI{1203.7}{\micro\second}} & \\
\midrule
& CRN-3 & 1985 & 8208  & 94.6\% & 100.0\% & 94.6\% & 97.2\% & CNN + Data Augmentation + DSC + Input Pooling  & {\SI{ 343.2}{\micro\second}} & \checkmark \\

\midrule[1pt]

{\cite{zeng2022cclsignal}}          & RF       & \cdash   & \cdash   & 99.7\% & 99.7\%  & 99.7\%  & 99.7\% & Random Forest                             & \cdash                    & \\
\midrule
{\cite{zeng2022cclsignal}}          & Adaboost & \cdash   & \cdash   & 99.6\% & 99.6\%  & 99.6\%  & 99.6\% & Adaboost                                  & \cdash                    & \\
\midrule
{\cite{zeng2022cclsignal}}          & XGBoost  & \cdash   & \cdash   & 99.6\% & 99.6\%  & 99.6\%  & 99.6\% & eXtreme Gradient Boosting                 & \cdash                    & \\
\midrule
{\cite{zeng2022cclsignal}}          & SVM      & \cdash   & \cdash   & 99.5\% & 99.5\%  & 99.5\%  & 99.5\% & Support Vector Machine                    & \cdash                    & \\
\midrule
{\cite{xiao2025dataaugmented}}      & TAN      & 21744160 & 31112608 & 97.7\% & 97.7\%  & 100.0\% & 98.9\% & Thin AlexNet + Data Augmentation          & \cdash                    & \\
\midrule
{\cite{xiao2025realization}}        & DTPP     & \cdash   & \cdash   & 97.3\% & 98.8\%  & 98.5\%  & 98.6\% & Dynamic Threshold + Physical Plausibility & {\SI{1.5}{\micro\second}} & \checkmark \\
\midrule
{\cite{raman2024data}}              & BiLSTM   & \cdash   & \cdash   & \cdash & 98.9\%  & 98.0\%  & 98.5\% & BiLSTM                                    & \cdash                    & \\
\midrule
{\cite{xiao2025dataaugmented}}      & MAN      & 1383584  & 3824640  & 96.2\% & 97.7\%  & 98.4\%  & 98.1\% & Miniaturized AlexNet + Data Augmentation  & \cdash                    & \\
\midrule
{\cite{jing2025identification}}     & CNN-LSTM & 65250    & 16086144 & 97.8\% & 95.9\%  & 99.1\%  & 97.5\% & CNN + LSTM                                & \cdash                    & \\
\midrule
{\cite{jing2025identification}}     & CNN      & \cdash   & \cdash   & 97.4\% & 100.0\% & 94.2\%  & 97.0\% & CNN                                       & \cdash                    & \\
\midrule
{\cite{raman2024data}}              & FNN      & \cdash   & \cdash   & \cdash & 93.6\%  & 97.9\%  & 95.7\% & Wavelet + FNN                             & \cdash                    & \\
\midrule
{\cite{raman2024data}}              & 1D-CNN   & \cdash   & \cdash   & \cdash & 94.0\%  & 96.3\%  & 95.1\% & CNN                                       & \cdash                    & \\
\midrule
{\cite{jing2025identification}}     & LSTM     & \cdash   & \cdash   & 94.8\% & 100.0\% & 88.4\%  & 93.9\% & LSTM                                      & \cdash                    & \\
\midrule
{\cite{torrescaceres2024automated}} & 1D-CNN   & 1814445  & 27451200 & \cdash & \cdash  & \cdash  & \cdash & CNN                                       & \cdash                    & \\

\bottomrule[1pt]
\end{tabular}
}
\vspace{0.5ex}
\noindent{\footnotesize\\
``--'' indicates ``Not reported'' or ``Not applicable''; ``Acc'', ``P'', ``R'', ``F1'' denote accuracy, precision, recall, and F1 score, respectively.
\\
Only models with reproducible architectural descriptions designed for CCL signal processing are listed.
\\
Note that the performance metrics of existing works are cited directly from their respective original papers; consequently, methodological paradigms and experimental conditions (e.g., datasets and evaluation methodologies) may vary.
}
\end{minipage}
\end{table*}

\let\cdash\undefined

\begin{figure*}[!b]
    \centering
    \includegraphics[width=.95\linewidth]{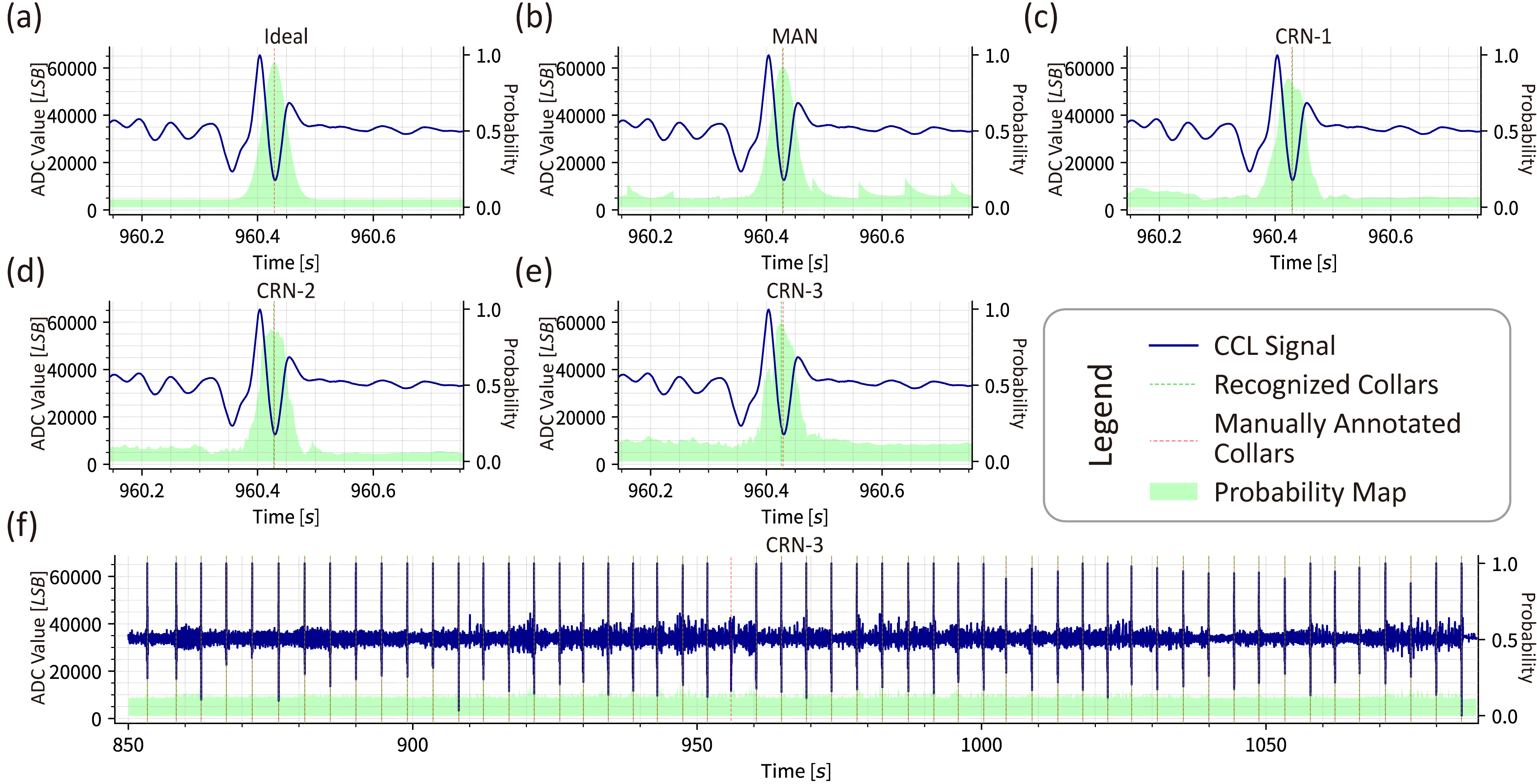}
    \caption{
        \textbf{(a)}~Ideal probability map derived from manual collar annotations;
        \textbf{(b)}~Probability map generated by MAN \cite{xiao2025dataaugmented};
        \textbf{(c)}~Probability map generated by CRN-1;
        \textbf{(d)}~Probability map generated by CRN-2;
        \textbf{(e)}~Probability map generated by CRN-3;
        Probability maps deviate more noticeably from the ideal map as network capacity decreases;
        \textbf{(f)}~An example of full-length recognition results using CRN-3, demonstrating that the majority of collar signatures are correctly recognized.
        \\
        For clarity, ``CCL Signal'' indicates the raw curve derived from CCL logs; ``Manually Annotated Collars'' indicates the centroid timestamps of manually annotated collar signatures, which serve as both training labels and evaluation ground truth; ``Recognized Collars'' indicates the centroid timestamps of collar signatures recognized by the proposed system; ``Probability Map'' represents the temporal curve of model output probabilities, representing the intermediate recognition result, where each time point indicates the probability of a collar signature at the current time step.
    }
    \label{fig4}
\end{figure*}

The proposed system is validated by evaluating both the neural network model and its MPU-based deployment. The model evaluation process involves performing inference on two full-length field CCL logs to recognize collar signatures. These logs comprise 55 and 77 collars, extending to maximum depth of \SI{512.2}{\meter} and \SI{771.1}{\meter}, respectively. To ensure an unbiased assessment, the logs for the evaluation were strictly excluded from the training set. The recognition results are compared with manually annotated ground truth collar labels to obtain model performance, including precision, recall and F1 scores. Specifically, recognized collar signatures within the temporal neighborhood of annotated collar labels are classified as true positives (TP); whereas detections outside these regions are classified as false positives (FP); and missed collar signatures are considered as false negatives (FN). The entire evaluation process is conducted offline on a workstation, as an example illustrated in Fig.~\ref{fig4}(f). The resulting performance for each model is detailed in Table~\ref{tab1} and~\ref{tab2}.

Results show that CRNs significantly reduce parameters and computational costs with only marginal performance loss. However, as network capacity decreases, probability maps deviate more noticeably from the ideal map, as illustrated in Fig.~\ref{fig4}(b)--(e). This trend highlights a trade-off between hardware efficiency and recognition precision. As shown in Table~\ref{tab2}, the proposed CRNs operate with only thousands of MACs, yet match or surpass the performance of existing models that demand millions of MACs.
Specifically, CRN-1 achieves an F1 score of 0.992, the highest among all evaluated models, while CRN-3 achieves an F1 score of 0.972. Compared with related recognition works, the computational cost of CRN-3 represents merely 264 ppm to 2140 ppm of the costs reported in \cite{xiao2025dataaugmented,jing2025identification,torrescaceres2024automated}. When benchmarked against standard TinyML architectures used in broader applications, such as MobileNetV3-small-0.75 (requiring 2.4M parameters and 44M MACs) \cite{howard2019searching} and the optimized DS-CNN-small (designed for keyword spotting (KWS) and requiring \textasciitilde 200K MACs) \cite{zhang2018helloedge}, the computational cost of CRN-3 is 186 ppm and 4.1\% of these models, respectively. It is important to note that a direct, unified comparison is challenging due to the diversity of methodological paradigms, the unavailability of algorithmic details, and the inaccessibility of private datasets in certain related works. Table~\ref{tab2} provides a broad overview of the performance landscape of existing works, where the performance metrics are cited directly from the original papers and may not be strictly comparable to the proposed models. Nevertheless, even considering variations in experimental setups, the proposed CRNs exhibit distinct advantages in performance, parameter efficiency, and computational economy relative to the existing literature.

The hardware implementation was validated by executing the CRN-3 model on the MPU using stored CCL waveforms. The end-to-end execution time, encompassing sample acquisition to final output, was recorded and compared against workstation-based simulations, as listed in Table~\ref{tab2}. The results indicate that the CRN-3 model inference requires an average of \SI{343.2}{\micro\second} per \SI{1}{\milli\second} sampling interval. This duration aligns with the computational capacity of the MPU and the computational complexity of the model, indicating the capability of the proposed solution to robustly recognize collar signatures in real-time while achieving a throughput of 1,000~IPS. The on-board probability maps and collar recognition results match the offline simulations within the margin of floating-point precision errors, thereby verifying the correctness of the embedded implementation.

Additionally, we estimate the power consumption of the CRN-3 model on the MPU. Based on the datasheet, the MPU power consumption is estimated to be \SI{197.2}{\milli\watt} under full load. However, when accounting for the inference duty cycle within each frame, the average power consumption of the CRN-3 model is approximately \SI{74.3}{\milli\watt}.

The dataset size in this study was constrained by the challenges associated with acquiring field data, which potentially limits the model's generalization capability. Nevertheless, the proposed system validates the feasibility of deploying lightweight neural networks for real-time, in-situ casing collar recognition. Future work will prioritize expanding the dataset by acquiring additional field data under more diverse conditions, such as extended RIH speed ranges, a larger number of wells, and more complex interference signals. Furthermore, integrating state-of-the-art advancements in neural network theory and signal processing is expected to yield higher recognition accuracy and lower computational costs, thereby further enhancing the performance of the recognition system.

\section{Conclusion}

This study presents a novel, lightweight casing collar recognition algorithm based on a domain-specific neural network architecture. By effectively balancing computational complexity with recognition performance, the most compact model variant, CRN-3, achieves a complexity of only 8,208 MACs while maintaining a high F1 score of 0.972. This represents a significant reduction in computational cost compared to standard TinyML baselines and existing works, with negligible degradation in performance.

The proposed approach overcomes the deployment barriers of existing collar recognition models caused by their prohibitive computational demands. To validate its practical applicability, the CRN-3 model was implemented on a casing collar recognition system powered by an ARM Cortex-M7 MPU for integration into downhole instruments. On-board validation confirms the system's capability to accurately recognize collar signatures under real-time constraints, achieving a processing throughput of 1,000~IPS.

This study demonstrates the feasibility of designing and deploying lightweight neural networks for real-time, in-situ casing collar recognition, which is a critical step toward the automation of downhole operations. By bridging the gap between advanced deep learning techniques and resource-constrained embedded environments within downhole instrumentation, this work paves the way for the development of fully autonomous downhole instrumentation.



\bibliographystyle{IEEEtran}
\bibliography{reference}

@inproceedings{ahmed2021case,
  title     = {Case Study -- Real Time Downhole Telemetry {CCL} and Tension Compression, a Differentiator for Successful Manipulation of {ICD's} in Horizontal Wells},
  author    = {Ahmed, Usman and Zhang, Zhiheng and Ortega Alfonzo, Ruben},
  booktitle = {SPE Middle East Oil \& Gas Show and Conference},
  _location = {Sanabis, Bahrain},
  pages     = {D031S026R006},
  year      = {2021},
  month     = {11},
  _ublisher = {SPE},
  _address  = {Richardson, TX, USA},
  doi       = {10.2118/204873-MS}
}

@article{alvarez2018theory,
  title     = {Theory, Design, Realization, and Field Results of An Inductive Casing Collar Locator},
  author    = {Alvarez, Jose Oliverio and Buzi, Erjola and Adams, Robert W and Deffenbaugh, Max},
  journal   = {IEEE Transactions on Instrumentation and Measurement},
  volume    = {67},
  number    = {4},
  pages     = {760--766},
  year      = {2018},
  _ublisher = {IEEE},
  _address  = {Piscataway, NJ, USA},
  doi       = {10.1109/TIM.2018.2795138}
}

@article{bai2018empirical,
  title         = {An Empirical Evaluation of Generic Convolutional and Recurrent Networks for Sequence Modeling},
  author        = {Bai, Shaojie and Kolter, J. Zico and Koltun, Vladlen},
  journal       = {arXiv preprint arXiv:1803.01271},
  year          = {2018},
  archiveprefix = {arXiv},
  primaryclass  = {cs.LG},
  doi           = {10.48550/arXiv.1803.01271}
}

@article{bouthillier2015dropout,
  title         = {Dropout as Data Augmentation},
  author        = {Bouthillier, Xavier and Konda, Kishore and Vincent, Pascal and Memisevic, Roland},
  journal       = {arXiv preprint arXiv:1506.08700},
  year          = {2015},
  archiveprefix = {arXiv},
  primaryclass  = {stat.ML},
  doi           = {10.48550/arXiv.1506.08700}
}

@article{brazell2019machine,
  title     = {A Machine-Learning-Based Approach to Assistive Well-Log Correlation},
  author    = {Brazell, Seth and Bayeh, Alex and Ashby, Michael and Burton, Darrin},
  journal   = {Petrophysics},
  volume    = {60},
  number    = {04},
  pages     = {469--479},
  year      = {2019},
  month     = {08},
  _ublisher = {SPWLA},
  _address  = {Houston, TX, USA},
  _issn     = {1529-9074},
  doi       = {10.30632/PJV60N4-2019a1}
}

@article{brown1990effects,
  title   = {The Effects of Cable on Signal Quality},
  author  = {Brown, Jim},
  journal = {Sound and Video Contractor},
  pages   = {22--33},
  year    = {1990}
}

@inproceedings{chollet2017xception,
  title     = {Xception: Deep Learning with Depthwise Separable Convolutions},
  author    = {Chollet, Fran{\c{c}}ois},
  booktitle = {IEEE Conference on Computer Vision and Pattern Recognition (CVPR)},
  _location = {Honolulu, HI, USA},
  pages     = {1251--1258},
  year      = {2017},
  month     = {07},
  _ublisher = {IEEE},
  _address  = {Piscataway, NJ, USA},
  doi       = {10.1109/CVPR.2017.195}
}

@article{cong2022perforating,
  title   = {Perforating Depth Control Method Based on Casing Hoop Automatic Tracking and Recognition Technology},
  author  = {Cong, Yan},
  journal = {Automation in Petro-Chemical Industry},
  volume  = {58},
  number  = {5},
  pages   = {29--33},
  year    = {2022},
  _issn   = {1007-7324}
}

@article{david2021tensorflow,
  title         = {TensorFlow Lite Micro: Embedded Machine Learning on TinyML Systems},
  author        = {Robert David and Jared Duke and Advait Jain and Vijay Janapa Reddi and Nat Jeffries and Jian Li and Nick Kreeger and Ian Nappier and Meghna Natraj and Shlomi Regev and Rocky Rhodes and Tiezhen Wang and Pete Warden},
  journal       = {arXiv preprint arXiv:2010.08678},
  year          = {2021},
  archiveprefix = {arXiv},
  primaryclass  = {cs.LG},
  doi           = {10.48550/arXiv.2010.08678}
}

@inproceedings{deffenbaugh2017untethered,
  title     = {An Untethered Sensor for Well Logging},
  author    = {Deffenbaugh, Max and Buzi, Erjola and Al-Maghrabi, Latifah and Ham, Gregory and Seren, Huseyin and Bernero, Greg},
  booktitle = {2017 IEEE Sensors Applications Symposium (SAS)},
  _location = {Glassboro, NJ, USA},
  year      = {2017},
  month     = {03},
  pages     = {1--5},
  _ublisher = {IEEE},
  _address  = {Piscataway, NJ, USA},
  doi       = {10.1109/SAS.2017.7894044}
}

@inproceedings{dosovitskiy2021an,
  title     = {An Image is Worth 16$\times$16 Words: Transformers for Image Recognition at Scale},
  author    = {Dosovitskiy, Alexey and Beyer, Lucas and Kolesnikov, Alexander and Weissenborn, Dirk and Zhai, Xiaohua and Unterthiner, Thomas and Dehghani, Mostafa and Minderer, Matthias and Heigold, Georg and Gelly, Sylvain and Uszkoreit, Jakob and Houlsby, Neil},
  booktitle = {International Conference on Learning Representations (ICLR)},
  _location = {Virtual Event, Austria},
  year      = {2021},
  month     = {05},
  _url      = {https://openreview.net/forum?id=YicbFdNTTy}
}

@inproceedings{elhadidy2025optimizing,
  title     = {Optimizing Well Perforation with Machine Learning: A Breakthrough in Predictive Modeling},
  author    = {Elhadidy, Ahmed and Helmy, Ahmed and Heikal, Mohamed and Hany, Wael},
  booktitle = {SPE Gas \& Oil Technology Showcase and Conference},
  _location = {Dubai, United Arab Emirates},
  pages     = {D022S002R002},
  year      = {2025},
  month     = {04},
  _ublisher = {SPE},
  _address  = {Richardson, TX, USA},
  doi       = {10.2118/224556-MS}
}

@inproceedings{entchev2011autonomous,
  title     = {Autonomous Perforating System for Multizone Completions},
  author    = {Entchev, Pavlin B. and Angeles, Renzo and Kumaran, Krishnan and Subrahmanya, Niranjan and Tolman, Randy},
  booktitle = {SPE Annual Technical Conference and Exhibition},
  _location = {Denver, CO, USA},
  year      = {2011},
  month     = {10},
  pages     = {SPE-147296-MS},
  _ublisher = {SPE},
  _address  = {Richardson, TX, USA},
  doi       = {10.2118/147296-MS}
}

@inproceedings{ghiasi2018dropblock,
  title     = {DropBlock: A Regularization Method for Convolutional Networks},
  author    = {Ghiasi, Golnaz and Lin, Tsung-Yi and Le, Quoc V.},
  booktitle = {Neural Information Processing Systems (NeurIPS)},
  _location = {Montreal, QC, Canada},
  year      = {2018},
  month     = {12},
  _ublisher = {Curran Associates, Inc.},
  _address  = {Red Hook, NY, USA},
  _url      = {https://proceedings.neurips.cc/paper/2018/file/7edcfb2d8f6a659ef4cd1e6c9b6d7079-Paper.pdf}
}

@inproceedings{gidado2023well,
  title     = {Well Diagnostic of New Underperforming Wells Using Downhole Log Tool {[SNT \& MDT]}},
  author    = {Gidado, A. O. and Ekesiobi, C. and Kpone-Tonwe, H. and Adesun, J.},
  booktitle = {SPE Nigeria Annual International Conference and Exhibition},
  _location = {Lagos, Nigeria},
  pages     = {D021S012R001},
  year      = {2023},
  month     = {07},
  _ublisher = {SPE},
  _address  = {Richardson, TX, USA},
  doi       = {10.2118/217236-MS}
}

@article{harris1966effect,
  title     = {The Effect of Perforating Oil Well Productivity},
  author    = {Harris, M.H.},
  journal   = {Journal of Petroleum Technology},
  volume    = {18},
  number    = {04},
  pages     = {518--528},
  year      = {1966},
  month     = {04},
  _ublisher = {SPE},
  _address  = {Richardson, TX, USA},
  _issn     = {0149-2136},
  doi       = {10.2118/1236-PA}
}

@article{howard2017mobilenets,
  title         = {{MobileNets}: Efficient Convolutional Neural Networks for Mobile Vision Applications},
  author        = {Howard, Andrew G. and Zhu, Menglong and Chen, Bo and Kalenichenko, Dmitry and Wang, Weijun and Weyand, Tobias and Andreetto, Marco and Adam, Hartwig},
  year          = {2017},
  journal       = {arXiv preprint arXiv:1704.04861},
  archiveprefix = {arXiv},
  primaryclass  = {cs.CV},
  doi           = {10.48550/arXiv.1704.04861}
}

@inproceedings{howard2019searching,
  title     = {Searching for {MobileNetV3}},
  author    = {Howard, Andrew and Sandler, Mark and Chu, Grace and Chen, Liang-Chieh and Chen, Bo and Tan, Mingxing and Wang, Weijun and Zhu, Yukun and Pang, Ruoming and Vasudevan, Vijay and Le, Quoc V. and Adam, Hartwig},
  booktitle = {IEEE/CVF International Conference on Computer Vision (ICCV)},
  _location = {Seoul, Republic of Korea},
  pages     = {1314--1324},
  year      = {2019},
  month     = {10},
  _ublisher = {IEEE},
  _address  = {Piscataway, NJ, USA},
  doi       = {10.1109/ICCV.2019.00140}
}

@inproceedings{hu2018squeeze,
  title     = {Squeeze-and-Excitation Networks},
  author    = {Hu, Jie and Shen, Li and Sun, Gang},
  booktitle = {IEEE Conference on Computer Vision and Pattern Recognition (CVPR)},
  _location = {Salt Lake City, UT, USA},
  pages     = {7132--7141},
  year      = {2018},
  month     = {06},
  _ublisher = {IEEE},
  _address  = {Piscataway, NJ, USA},
  doi       = {10.1109/CVPR.2018.00745}
}

@inproceedings{ioffe2015batch,
  title     = {Batch Normalization: Accelerating Deep Network Training by Reducing Internal Covariate Shift},
  author    = {Ioffe, Sergey and Szegedy, Christian},
  booktitle = {32nd International Conference on Machine Learning (ICML)},
  _location = {Lille, France},
  pages     = {448--456},
  year      = {2015},
  month     = {07},
  _ublisher = {PMLR},
  _address  = {Cambridge, MA, USA},
  _url      = {http://proceedings.mlr.press/v37/ioffe15.html}
}

@article{jing2025identification,
  title     = {Identification and Prediction of Casing Collar Signal Based on {CNN-LSTM}},
  author    = {Jing, Jun and Qin, Yiman and Zhu, Xiaohua and Shan, Hongbin and Peng, Peng},
  journal   = {Arabian Journal for Science and Engineering},
  volume    = {50},
  number    = {7},
  pages     = {4897--4911},
  year      = {2025},
  _ublisher = {Springer},
  _address  = {Berlin, Germany},
  _issn     = {2191-4281},
  doi       = {10.1007/s13369-024-09440-5}
}

@inproceedings{kim2021broadcasted,
  title     = {Broadcasted Residual Learning for Efficient Keyword Spotting},
  author    = {Kim, Byeonggeun and Chang, Simyung and Lee, Jinkyu and Sung, Dooyong},
  booktitle = {Interspeech 2021},
  _location = {Brno, Czechia},
  year      = {2021},
  month     = {08},
  pages     = {4538--4542},
  _issn     = {2958-1796},
  doi       = {10.21437/Interspeech.2021-383}
}

@article{li2010approach,
  title   = {Study of the Approach to Extract Casing Collar Locator Information Features Based on Anti-aliasing Wavelet Time Entropy in Frequency Domain},
  author  = {Li, Haoyu and Chen, Jikai and Xiao, Yong and Liu, Xingbin and Wu, Jianqiang},
  journal = {High Technology Letters},
  year    = {2010},
  volume  = {20},
  number  = {5},
  pages   = {538--543},
  _issn   = {1002-0470},
  doi     = {10.3772/j._issn1002-0470.2010.05.016}
}

@inproceedings{li2013casing,
  title     = {Casing State Detection Methods Based on the {CCL} Signal of the Tractor for Horizontal Wells},
  author    = {Li, Haoyu and Tang, Tiantian and Wang, Yanjun},
  booktitle = {2013 IEEE 11th International Conference on Electronic Measurement \& Instruments},
  _location = {Harbin, China},
  volume    = {2},
  pages     = {568--573},
  year      = {2013},
  month     = {08},
  _ublisher = {IEEE},
  _address  = {Piscataway, NJ, USA},
  doi       = {10.1109/ICEMI.2013.6743143}
}

@article{li2020application,
  title   = {Application of Cross Correlation Function Method in Locating Perforation Depth},
  author  = {Li, Jin and Liu, Yuhai and Zhang, Jian and Wang, Jiang and Zhang, Yiling},
  journal = {Journal of Southwest Petroleum University (Natural Science Edition)},
  year    = {2020},
  volume  = {42},
  number  = {6},
  pages   = {42--48},
  _issn   = {1674-5086},
  doi     = {10.11885/j._issn1674-5086.2020.06.24.01}
}

@inproceedings{lin2013network,
  title     = {Network In Network},
  author    = {Lin, Min and Chen, Qiang and Yan, Shuicheng},
  booktitle = {International Conference on Learning Representations (ICLR)},
  _location = {Banff, AB, Canada},
  year      = {2014},
  month     = {04}
}

@inproceedings{lin2020mcunet,
  title     = {MCUNet: Tiny Deep Learning on IoT Devices},
  author    = {Lin, Ji and Chen, Wei-Ming and Lin, Yujun and cohn, john and Gan, Chuang and Han, Song},
  booktitle = {Neural Information Processing Systems (NeurIPS)},
  _location = {Online},
  year      = {2020},
  month     = {12},
  volume    = {33},
  pages     = {11711--11722},
  _ublisher = {Curran Associates, Inc.},
  _address  = {Red Hook, NY, USA},
  _url      = {https://proceedings.neurips.cc/paper_files/paper/2020/file/86c51678350f656dcc7f490a43946ee5-Paper.pdf}
}

@inproceedings{lin2021mcunetv2,
  title     = {MCUNetV2: Memory-Efficient Patch-based Inference for Tiny Deep Learning},
  author    = {Lin, Ji and Chen, Wei-Ming and Cai, Han and Gan, Chuang and Han, Song},
  booktitle = {Neural Information Processing Systems (NeurIPS)},
  _location = {Online},
  year      = {2021},
  volume    = {34},
  pages     = {2346--2358},
  _ublisher = {Curran Associates, Inc.},
  _address  = {Red Hook, NY, USA},
  _url      = {https://proceedings.neurips.cc/paper_files/paper/2021/file/1371bccec2447b5aa6d96d2a540fb401-Paper.pdf}
}

@inproceedings{lin2022ondevice,
  title     = {On-Device Training Under 256KB Memory},
  author    = {Lin, Ji and Zhu, Ligeng and Chen, Wei-Ming and Wang, Wei-Chen and Gan, Chuang and Han, Song},
  booktitle = {Neural Information Processing Systems (NeurIPS)},
  _location = {New Orleans, LA, USA},
  year      = {2022},
  month     = {11},
  volume    = {35},
  pages     = {22941--22954},
  _ublisher = {Curran Associates, Inc.},
  _address  = {Red Hook, NY, USA},
  _url      = {https://proceedings.neurips.cc/paper_files/paper/2022/file/90c56c77c6df45fc8e556a096b7a2b2e-Paper-Conference.pdf}
}

@article{liu2025research,
  title     = {Research on Well Depth Tracking Calculation Method Based on Branching Parallel Neural Networks},
  author    = {Liu, Weikai and Ma, Baoquan and Yu, Xiaolei},
  journal   = {Processes},
  volume    = {13},
  year      = {2025},
  number    = {10},
  pages     = {3147},
  _ublisher = {MDPI},
  _address  = {Basel, Switzerland},
  _issn     = {2227-9717},
  doi       = {10.3390/pr13103147}
}

@book{lu2012oilandgasfield,
  title     = {Oil \& Gas Field Perforating Technology},
  author    = {Lu, Dawei},
  publisher = {Petroleum Industry Press},
  address   = {Beijing, China},
  year      = {2012},
  isbn      = {978-7-5021-9026-2}
}

@article{lu2025digital,
  title     = {Digital twin-driven water-wave information transmission and recurrent acceleration network for remaining useful life prediction of gear box},
  author    = {Lu, Quanbo and Huang, Xiaojuan and Wu, Guangjie and Shen, Xinqi and Zhu, Dong},
  journal   = {Engineering Research Express},
  volume    = {7},
  number    = {2},
  pages     = {025202},
  year      = {2025},
  month     = {4},
  _ublisher = {IOP Publishing},
  doi       = {10.1088/2631-8695/adc076}
}

@inproceedings{mijarez2014hpht,
  title     = {{HPHT} Cased-Hole {CCL} Tool Enhancement via {DSP} Techniques for Accurate Depth Control in Wire-Line Well Interventions},
  author    = {Mijarez, Rito and Pascacio, David and Guevara, Ricardo and Tello, Carlos and Pacheco, Olimpia and Rodr{\'\i}guez, Joaqu{\'\i}n},
  booktitle = {International Conference on High Temperature Electronics},
  _location = {Albuquerque, NM, USA},
  pages     = {305--310},
  year      = {2014},
  month     = {05},
  _ublisher = {IMAPS},
  _address  = {Pittsburgh, PA, USA},
  doi       = {10.4071/HITEC-THA15}
}

@article{noh2021deep,
  title         = {Deep-Learning Inversion Method for the Interpretation of Noisy Logging-While-Drilling Resistivity Measurements},
  author        = {Kyubo Noh and David Pardo and Carlos Torres-Verdin},
  journal       = {arXiv preprint arXiv:2111.07490},
  year          = {2021},
  archiveprefix = {arXiv},
  primaryclass  = {physics.geo-ph},
  doi           = {10.48550/arXiv.2111.07490}
}

@article{oord2016wavenet,
  title         = {WaveNet: A Generative Model for Raw Audio},
  author        = {van den Oord, Aaron and Dieleman, Sander and Zen, Heiga and Simonyan, Karen and Vinyals, Oriol and Graves, Alex and Kalchbrenner, Nal and Senior, Andrew and Kavukcuoglu, Koray},
  journal       = {arXiv preprint arXiv:1609.03499},
  year          = {2016},
  archiveprefix = {arXiv},
  primaryclass  = {cs.SD},
  doi           = {10.48550/arXiv.1609.03499}
}

@inproceedings{raman2024data,
  title     = {Data Driven Casing Collar Feature Detection and Identification for Automated Depth Estimation for Wireline},
  author    = {Raman, S. K. and Abuhaikal, M.},
  booktitle = {Fourth EAGE Digitalization Conference \& Exhibition, },
  _location = {Paris, France},
  pages     = {1--5},
  year      = {2024},
  month     = {03},
  _ublisher = {EAGE},
  _address  = {Utrecht, The Netherlands},
  _issn     = {2214-4609},
  doi       = {10.3997/2214-4609.202439084}
}

@inproceedings{sandler2018mobilenetv2,
  title     = {{MobileNetV2}: Inverted Residuals and Linear Bottlenecks},
  author    = {Sandler, Mark and Howard, Andrew and Zhu, Menglong and Zhmoginov, Andrey and Chen, Liang-Chieh},
  booktitle = {IEEE Conference on Computer Vision and Pattern Recognition (CVPR)},
  _location = {Salt Lake City, UT, USA},
  pages     = {4510--4520},
  year      = {2018},
  _ublisher = {IEEE},
  _address  = {Piscataway, NJ, USA},
  doi       = {10.1109/CVPR.2018.00474}
}

@article{seren2022miniaturized,
  title     = {Miniaturized Casing Collar Locator for Small Downhole Robots},
  author    = {Seren, Huseyin Rahmi and Deffenbaugh, Max},
  journal   = {IEEE Sensors Letters},
  volume    = {6},
  number    = {4},
  pages     = {1--4},
  year      = {2022},
  _ublisher = {IEEE},
  _address  = {Piscataway, NJ, USA},
  doi       = {10.1109/LSENS.2022.3158002}
}

@inproceedings{seren2023magnetic,
  title     = {Magnetic Sensors for Navigation of Untethered Downhole Robots},
  author    = {Seren, Huseyin and Deffenbaugh, Max and Larbi Zeghlache, Mohamed and Bukhamseen, Ahmed},
  booktitle = {SPE Middle East Oil and Gas Show and Conference},
  _location = {Manama, Bahrain},
  pages     = {D021S087R004},
  year      = {2023},
  month     = {03},
  _ublisher = {SPE},
  _address  = {Richardson, TX, USA},
  doi       = {10.2118/213358-MS}
}

@inproceedings{seren2025probabilistic,
  title     = {Probabilistic Casing Collar Locator for Untethered Downhole Tools},
  author    = {Seren, Huseyin and Zeghlache, Mohamed Larbi and Deffenbaugh, Max and Jabari, Rami},
  booktitle = {Abu Dhabi International Petroleum Exhibition and Conference},
  _location = {Abu Dhabi, United Arab Emirates},
  pages     = {D031S091R006},
  year      = {2025},
  month     = {11},
  _ublisher = {SPE},
  _address  = {Richardson, TX, USA},
  doi       = {10.2118/229500-MS}
}

@article{simonyan2015very,
  title         = {Very Deep Convolutional Networks for Large-Scale Image Recognition},
  author        = {Simonyan, Karen and Zisserman, Andrew},
  journal       = {arXiv preprint arXiv:1409.1556},
  year          = {2015},
  archiveprefix = {arXiv},
  primaryclass  = {cs.CV},
  doi           = {10.48550/arXiv.1409.1556}
}

@article{song2023finite,
  title   = {Finite Element Analysis on Responses of {CCL} Log},
  author  = {Song, Jiajia and Ji, Yongli and Xia, Jigen},
  journal = {Petroleum Tubular Goods \& Instruments},
  volume  = {9},
  number  = {02},
  pages   = {66--70},
  year    = {2023},
  _issn   = {2096-0077},
  doi     = {10.19459/j.cnki.61-1500/te.2023.02.014}
}

@mastersthesis{srivastava2013improving,
  title  = {Improving Neural Networks with Dropout},
  author = {Srivastava, Nitish},
  school = {University of Toronto},
  year   = {2013},
  _url   = {https://www.cs.toronto.edu/~nitish/msc_thesis.pdf}
}

@article{srivastava2014dropout,
  title   = {Dropout: A Simple Way to Prevent Neural Networks from Overfitting},
  author  = {Srivastava, Nitish and Hinton, Geoffrey and Krizhevsky, Alex and Sutskever, Ilya and Salakhutdinov, Ruslan},
  journal = {Journal of Machine Learning Research},
  volume  = {15},
  number  = {1},
  pages   = {1929--1958},
  year    = {2014},
  _url    = {http://jmlr.org/papers/v15/srivastava14a.html}
}

@inproceedings{szegedy2016rethinking,
  title     = {Rethinking the Inception Architecture for Computer Vision},
  author    = {Szegedy, Christian and Vanhoucke, Vincent and Ioffe, Sergey and Shlens, Jon and Wojna, Zbigniew},
  booktitle = {IEEE Conference on Computer Vision and Pattern Recognition (CVPR)},
  _location = {Las Vegas, NV, USA},
  pages     = {2818--2826},
  year      = {2016},
  month     = {06},
  _ublisher = {IEEE},
  _address  = {Piscataway, NJ, USA},
  doi       = {10.1109/CVPR.2016.308}
}

@inproceedings{tompson2015efficient,
  title     = {Efficient Object Localization Using Convolutional Networks},
  author    = {Tompson, Jonathan and Goroshin, Ross and Jain, Arjun and LeCun, Yann and Bregler, Christoph},
  booktitle = {IEEE Conference on Computer Vision and Pattern Recognition (CVPR)},
  _location = {Boston, MA, USA},
  pages     = {648--656},
  year      = {2015},
  month     = {06},
  _ublisher = {IEEE},
  _address  = {Piscataway, NJ, USA},
  doi       = {10.1109/CVPR.2015.7298664}
}

@article{torrescaceres2024automated,
  title     = {Automated Well Log Depth Matching: Late Fusion Multimodal Deep Learning},
  author    = {Torres Caceres, Veronica Alejandra and Duffaut, Kenneth and Yazidi, Anis and Westad, Frank and Johansen, Yngve Bolstad},
  journal   = {Geophysical Prospecting},
  volume    = {72},
  number    = {1},
  pages     = {155--182},
  year      = {2024},
  _ublisher = {EAGE},
  _address  = {Amsterdam, The Netherlands},
  _issn     = {1365-2478},
  doi       = {10.1111/1365-2478.13200}
}

@inproceedings{van2016conditional,
  title     = {Conditional Image Generation with PixelCNN Decoders},
  author    = {van den Oord, Aaron and Kalchbrenner, Nal and Espeholt, Lasse and Vinyals, Oriol and Graves, Alex},
  booktitle = {Neural Information Processing Systems (NeurIPS)},
  _location = {Barcelona, Spain},
  year      = {2016},
  month     = {12},
  _ublisher = {Curran Associates, Inc.},
  _address  = {Red Hook, NY, USA},
  _url      = {https://proceedings.neurips.cc/paper/2016/file/b1301141feffabac455e1f90a7de2054-Paper.pdf}
}

@inproceedings{vaswani2017attention,
  title     = {Attention Is All You Need},
  author    = {Vaswani, Ashish and Shazeer, Noam and Parmar, Niki and Uszkoreit, Jakob and Jones, Llion and Gomez, Aidan N. and Kaiser, {\L}ukasz and Polosukhin, Illia},
  booktitle = {Neural Information Processing Systems (NeurIPS)},
  _location = {Long Beach, CA, USA},
  year      = {2017},
  month     = {12},
  _ublisher = {Curran Associates, Inc.},
  _address  = {Red Hook, NY, USA},
  _url      = {https://proceedings.neurips.cc/paper/2017/file/3f5ee243547dee91fbd053c1c4a845aa-Paper.pdf}
}

@article{viggen2025improving,
  title     = {Improving Pipe Perforation Estimates from Ultrasonic Imaging Using Subpixel Machine Learning Trained on Optical Data},
  author    = {Viggen, Erlend Magnus and Gr{\o}nsberg, Sondre and Brekke, Svein and Hicks, Brad and Wifstad, Sigurd Vangen},
  journal   = {Geoenergy Science and Engineering},
  volume    = {246},
  pages     = {213541},
  year      = {2025},
  _ublisher = {Elsevier},
  _address  = {Amsterdam, The Netherlands},
  _issn     = {2949-8910},
  doi       = {10.1016/j.geoen.2024.213541}
}

@article{wang2006application,
  title   = {Application of Computer Automatic Discriminating Technology to the Depth Control of Perforation},
  author  = {Wang, Hong-Liang and Tang, Wen-Jiang},
  journal = {Well Logging Technology},
  volume  = {30},
  number  = {4},
  pages   = {378--380},
  year    = {2006},
  _issn   = {1004-1338},
  doi     = {10.16489/j._issn1004-1338.2006.04.027}
}

@article{wang2012collardepth,
  title   = {Study on Collar Depth Identification Based on Relative Amplitude Method},
  author  = {Wang, Hui and Lv, Haixia and Pan, Junhui and Li, Guojia and Gao, Xing},
  journal = {Journal of Harbin University of Commerce (Natural Sciences Edition)},
  volume  = {28},
  number  = {4},
  pages   = {435--438},
  year    = {2012},
  _issn   = {1672-0946},
  doi     = {10.19492/j.cnki.1672-0946.2012.04.016}
}

@inproceedings{wang2017timeseries,
  title     = {Time series classification from scratch with deep neural networks: A strong baseline},
  author    = {Wang, Zhiguang and Yan, Weizhong and Oates, Tim},
  booktitle = {2017 International Joint Conference on Neural Networks (IJCNN)},
  _location = {Anchorage, AK, USA},
  year      = {2017},
  month     = {05},
  pages     = {1578-1585},
  doi       = {10.1109/IJCNN.2017.7966039}
}

@article{xiao2025dataaugmented,
  title          = {Data-Augmented Deep Learning for Downhole Depth Sensing and Validation},
  author         = {Xiao, Si-Yu and Zhao, Xin-Di and Mao, Tian-Hao and Wang, Yi-Wei and Chen, Yu-Qiao and Zhang, Hong-Yun and Wang, Jian and Wang, Jun-Jie and Liu, Shuang and Chen, Tu-Pei and Liu, Yang},
  journal        = {Sensors},
  volume         = {26},
  year           = {2026},
  number         = {3},
  article-number = {775},
  pubmedid       = {41682291},
  _issn          = {1424-8220},
  _ublisher      = {MDPI},
  _address       = {Basel, Switzerland},
  doi            = {10.3390/s26030775}
}

@article{xiao2025realization,
  title         = {Realization of Precise Perforating Using Dynamic Threshold and Physical Plausibility Algorithm for Self-Locating Perforating in Oil and Gas Wells},
  author        = {Xiao, Si-Yu and Ren, Guo-Hui and Mao, Tian-Hao and Chen, Yu-Qiao and Liu, Yi-An and Wang, Jun-Jie and Tang, Kai and Zhao, Xin-Di and Yu, Zhi-Jian and Liu, Shuang and Chen, Tu-Pei and Yang, Liu},
  journal       = {arXiv preprint arXiv:2509.00608},
  year          = {2025},
  archiveprefix = {arXiv},
  primaryclass  = {eess.SY},
  doi           = {10.48550/arXiv.2509.00608}
}

@article{yan2024automatic,
  title   = {Automatic Identification Method of Collar Based on {Faster-RCNN} Network},
  author  = {Yan, Zhengguo and Chen, Ying and Zou, Shijiao and Li, Jinjiang},
  journal = {Industrial Control Computer},
  volume  = {37},
  number  = {3},
  pages   = {57--58},
  year    = {2024}
}

@article{yang2025leak,
  title     = {Leak Identification and Positioning Strategies for Downhole Tubing in Gas Wells},
  author    = {Yang, Yun Peng and Luan, Guo Hua and Zhang, Lian Fang and Niu, Ming Yong and Zou, Guang Gui and Zhang, Xu Liang and Wang, Jin You and Yang, Jing Feng and Li, Mo Song},
  journal   = {Processes},
  volume    = {13},
  number    = {6},
  pages     = {1708},
  year      = {2025},
  _ublisher = {MDPI},
  _address  = {Basel, Switzerland},
  _issn     = {2227-9717},
  doi       = {10.3390/pr13061708}
}

@inproceedings{zeng2022cclsignal,
  title     = {Research on {CCL} Signal Recognition Method for Coupling Based on {SVM} Algorithm},
  author    = {Zeng, Yuting},
  booktitle = {2022 7th International Conference on Cloud Computing and Big Data Analytics (ICCCBDA)},
  _location = {Chengdu, China},
  year      = {2022},
  month     = {04},
  pages     = {331-336},
  _ublisher = {IEEE},
  _address  = {Piscataway, NJ, USA},
  doi       = {10.1109/ICCCBDA55098.2022.9778871}
}

@article{zhang2018helloedge,
  title         = {Hello Edge: Keyword Spotting on Microcontrollers},
  author        = {Zhang, Yundong and Suda, Naveen and Lai, Liangzhen and Chandra, Vikas},
  journal       = {arXiv preprint arXiv:1711.07128},
  year          = {2018},
  archiveprefix = {arXiv},
  primaryclass  = {cs.SD},
  doi           = {10.48550/arXiv.1711.07128}
}

@article{zhang2024yolo,
  title   = {{YOLOv5}-Based Detection Method for Oil and Gas Well Casing Joints},
  author  = {Zhang, Jiatian and Zhao, Yao and Yan, Zhengguo and Ren, Xing and Zhang, Zhiwei},
  journal = {Journal of Xi'an Shiyou University (Natural Science Edition)},
  volume  = {39},
  number  = {04},
  pages   = {83--89},
  year    = {2024},
  _issn   = {1673-064X}
}

@article{zhao2021highspeed,
  title     = {A High-Speed Well Logging Telemetry System Based on Low-Power {FPGA}},
  author    = {Zhao, Hongwei and Song, Kezhu and Li, Kehan and Wu, Chuan and Chen, Zhuo},
  journal   = {IEEE Access},
  year      = {2021},
  volume    = {9},
  pages     = {8178--8191},
  _ublisher = {IEEE},
  _address  = {Piscataway, NJ, USA},
  _issn     = {2169-3536},
  doi       = {10.1109/ACCESS.2021.3049799}
}

@inproceedings{zhao2022detection,
  title     = {Detection Method of Casing Joint Based on Computer Vision},
  author    = {Zhao, Yao and Zhang, Jiatian and Guo, Liang and Zhang, Zhiwei},
  booktitle = {2022 4th International Conference on Intelligent Control, Measurement and Signal Processing (ICMSP)},
  _location = {Hangzhou, China},
  year      = {2022},
  month     = {07},
  pages     = {1006--1009},
  _ublisher = {IEEE},
  _address  = {Piscataway, NJ, USA},
  doi       = {10.1109/ICMSP55950.2022.9859086}
}

\vfill

\section*{Biographies}

\begin{IEEEbiography}[{\includegraphics[width=1in,height=1.25in,clip,keepaspectratio]{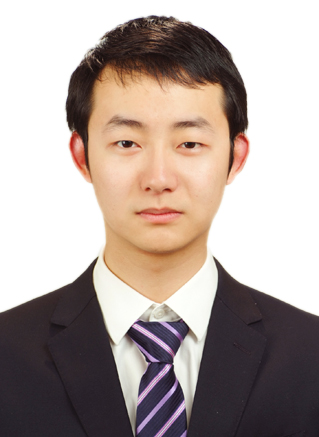}}]
~~
Si-Yu Xiao received the B.S. degree~in Microelectronic Science and Engineering from University of Electronic Science and Technology of China (UESTC) in 2018. He is currently pursuing the Ph.D. degree in Electronic Science and Technology. His research interests include digital circuit design, edge computing, artificial intelligence and their applications.
\end{IEEEbiography}

\begin{IEEEbiography}[{\includegraphics[width=1in,height=1.25in,clip,keepaspectratio]{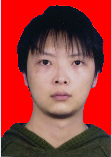}}]
~
Xin-Di Zhao received the Bachelor of Engineering degree in Software Engineering from Southwest Petroleum University. He is now a Level 2 Engineer in perforation technology research at the Southwest Branch of China National Petroleum Corporation Logging Co., Ltd. His current research includes new perforation tools, technologies and methods.
\end{IEEEbiography}

\begin{IEEEbiography}[{\includegraphics[width=1in,height=1.25in,clip,keepaspectratio]{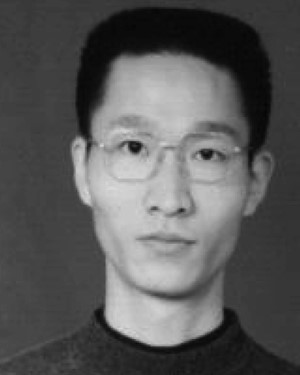}}]
~
Xiang-Zhan Wang received the Ph.D. degree in microelectronics and solid state electronics from the University of Electronic Science and Technology of China (UESTC), in 2010. He is currently with the State Key Laboratory of Electronic Thin Films and Integrated Devices, UESTC.
\end{IEEEbiography}

\begin{IEEEbiography}[{\includegraphics[width=1in,height=1.25in,clip,keepaspectratio]{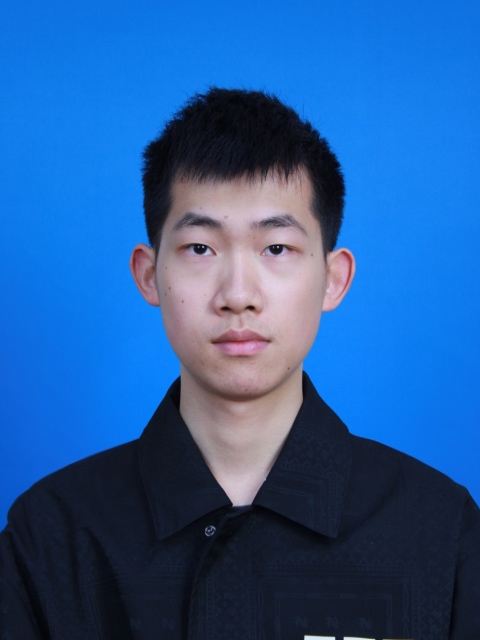}}]
~
Tian-Hao Mao is currently pursuing a Bachelor's degree in Integrated Circuit Design and Integrated Systems at the University of Electronic Science and Technology of China. His academic interests focus on digital circuit design, system-on-chip (SoC) architecture, and algorithm optimization for hardware acceleration.
\end{IEEEbiography}

\begin{IEEEbiography}[{\includegraphics[width=1in,height=1.25in,clip,keepaspectratio]{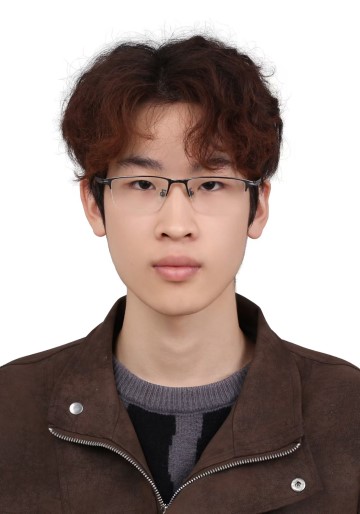}}]
~
Ying-Kai Liao is currently pursuing a bachelor's degree in Microelectronic Engineering at the University of Electronic Science and Technology of China (UESTC), Chengdu, China.
\end{IEEEbiography}

\begin{IEEEbiography}[{\includegraphics[width=1in,height=1.25in,clip,keepaspectratio]{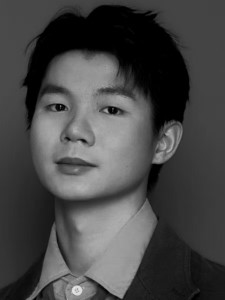}}]
~
Xing-Yu Liao obtained a bachelor's degree in Microelectronics Science and Engineering from the University of Electronic Science and Technology of China (UESTC) in 2024. He is currently pursuing a doctoral degree in the School of Integrated Circuit Science and Engineering at UESTC. His research interests include digital integrated circuit design, AI accelerator architecture and embedded system design.
\end{IEEEbiography}

\begin{IEEEbiography}[{\includegraphics[width=1in,height=1.25in,clip,keepaspectratio]{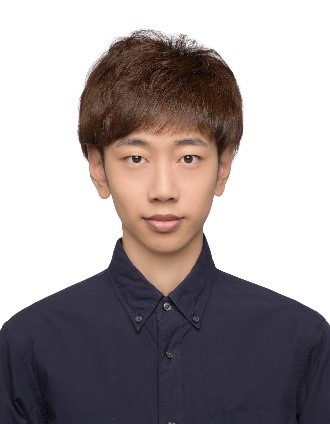}}]
~
Yu-Qiao Chen is currently pursuing the M.S. degree in Integrated Circuit Science and Engineering at the University of Electronic Science and Technology of China, where he also received the B.S. degree in Microelectronic Science and Engineering. His research interests cover digital circuit design, System-on-Chip (SoC), and neural network algorithms.
\end{IEEEbiography}

\begin{IEEEbiography}[{\includegraphics[width=1in,height=1.25in,clip,keepaspectratio]{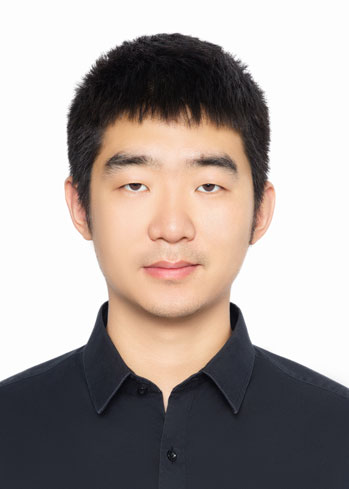}}]
~
Jun-Jie Wang received the Ph.D. degree in microelectronics from the University of Electronic Science and Technology of China. He is now a researcher at the University of Electronic Science and Technology of China. His current research interests include digital circuit design, nonvolatile memory devices, and their applications in artificial intelligence.
\end{IEEEbiography}

\begin{IEEEbiography}[{\includegraphics[width=1in,height=1.25in,clip,keepaspectratio]{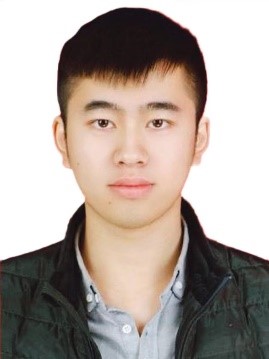}}]
~
Shuang Liu received the Ph.D. degree in microelectronics from the University of Electronic Science and Technology of China. He is now a postdoctoral researcher at the University of Electronic Science and Technology of China. His current research interests include processing{-}in{-}memory circuits and neuromorphic systems.
\end{IEEEbiography}

\begin{IEEEbiography}[{\includegraphics[width=1in,height=1.25in,clip,keepaspectratio]{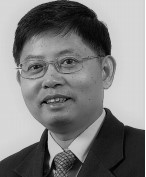}}]
~
Tu-Pei Chen is a tenured faculty member in the School of Electrical and Electronic Engineering at Nanyang Technological University, Singapore, where he has been teaching and conducting research in the field of Microelectronics for over 25 years. His current research interests include on-chip ESD and latch-up protection, memory devices, memory-based computing (in-memory computing, neuromorphic computing), and thin-film transistors and applications.
\end{IEEEbiography}

\begin{IEEEbiography}[{\includegraphics[width=1in,height=1.25in,clip,keepaspectratio]{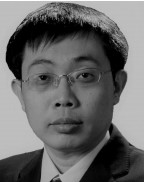}}]
~
Yang Liu received the B.Sc. degree in microelectronics from Jilin University, China, in 1998 and the Ph.D. degree from Nanyang Technological University, Singapore, in 2005. From May 2005 to July 2006, he was a Research Fellow with Nanyang Technological University, Singapore. In 2006, he was awarded the prestigious Singapore Millennium Foundation Fellowship. In 2008, he joined the School of Microelectronics, University of Electronic Science and Technology, China, as a full professor. He is the author or coauthor of over 130 peer-reviewed journal papers and more than 100 conference papers. He has been awarded one US patent and more than 30 China patents also. His current research includes memristor neural network system, neuromorphic computing ICs, and AI-RFICs.
\end{IEEEbiography}

\vfill

\end{document}